\documentclass[conference]{IEEEtran}
\IEEEoverridecommandlockouts
\usepackage{cite}
\usepackage{amsmath,amssymb,amsfonts}
\usepackage{algorithmic}
\usepackage{graphicx}
\usepackage{subcaption}
\usepackage{textcomp}
\usepackage{xcolor}
\usepackage{gensymb}

\def\BibTeX{{\rm B\kern-.05em{\sc i\kern-.025em b}\kern-.08em
    T\kern-.1667em\lower.7ex\hbox{E}\kern-.125emX}}
    
\begin{document}

\title{Active Noise Control for Augmented Reality Glasses using Virtual Sensing with Directional Robustness \\
}

\author{%
Nitay Peled, Boaz Rafaely\\[0.5em]
\textit{School of Electrical and Computer Engineering}\\
\textit{Ben-Gurion University of the Negev}\\
Beer-Sheva, Israel\\
Email: nitaype@post.bgu.ac.il, br@bgu.ac.il
}
\maketitle

\begin{abstract}
Active noise control (ANC) is an established technique for mitigating acoustic noise, successfully implemented in various commercial applications. However, the application of ANC to augmented reality (AR) glasses remains an emerging area, where a key challenge is the lack of a microphone at the user's ear. This paper first validates the necessity of virtual sensing techniques for this specific geometry and investigates the Auxiliary Filter (AF) method as a potential solution. We present a theoretical analysis of the method's directional sensitivity, enabling the prediction of the system's robustness when the noise direction deviates from the calibration angle. Simulation results confirm that while a fixed AF is effective under matched conditions, it suffers significant degradation under mismatch, consistent with the theoretical predictions. To address this limitation, we propose and evaluate a solution using a finite bank of auxiliary filters that are selected based on the noise source direction of arrival. The results demonstrate that this filter bank approach significantly recovers performance compared to a fixed filter, offering a practical alternative to continuous recalibration in dynamic acoustic environments.
\end{abstract}


\section{Introduction}

Active noise control (ANC) is an established technique for addressing the problem of everyday acoustic noise, in which a loudspeaker cancels unwanted sound by generating anti-noise of equal magnitude and opposite phase \cite{kuo_active_1999, nelson_active_1991}. Over the past few decades, significant advancements have been made in ANC algorithms, which have been successfully implemented in various practical applications \cite{lu_survey_2021}. These include headphones, headrests, and more \cite{chang_listening_2016}. While well established in some domains, the application of ANC to augmented reality (AR) glasses remains an emerging field of research.

ANC systems utilize microphones to acquire error and optionally reference signals, then process these signals to finally drive loudspeakers that aim to attenuate noise at a desired location \cite{ANC_book}. AR glasses can include microphones and loudspeakers around the glasses’ frame, making them applicable to ANC. However, they typically do not include a microphone at the ear, as may be desired. Furthermore, the limitation of having the microphones fixed on the frame of the glasses introduces a causality issue as it allows noise to arrive from directions where the sound wave reaches the ear before reaching the microphone. 

Several recent works have investigated ANC systems related to AR-glasses. The study in \cite{xiao_spatially_2023} demonstrated directional noise reduction in an AR-glasses simulation but relied on binaural error microphones, which may not be available in practice. Other studies \cite{miyake_head-mounted_2023, miyazaki_head-mounted_2015} examined head-mounted arrangements with certain similarities to AR glasses; however, their microphone placements and operating conditions—such as using an error microphone very close to the ear or focusing on magnetic resonance noise scenarios—differ from those feasible in actual AR-glasses designs.

While solutions to the problem of the absence of an error microphone at the ear was not studied specifically for AR glasses, it was studied quite comprehensively in the general case. This limitation is typically addressed by virtual sensing (VS) techniques. VS methods enable ANC systems to attenuate sound pressure at locations where no physical microphones are present \cite{liu_virtual_2009}. Several recent studies have analyzed and compared different VS approaches, including the Auxiliary Filter (AF) method \cite{pawelczyk_adaptive_2004, zhang_robust_2021, shi_active_2020, kajikawa_comparison_2019}, which is the method considered here. 
Although it is a promising solution, the method requires calibration, or offline training, and its performance may significantly degrade with mismatch in conditions between training and control, necessitating re-training \cite{chuang_shi_selective_2019, xie_cognitive_2024}. Consequently, several techniques for online adaptation of the AF have been proposed in recent years \cite{chuang_shi_selective_2019, xie_cognitive_2024, wang_transferable_2025}. Contemporary approaches rely on a bank of trained AFs to cover a range of conditions. While promising, they have not been studied previously for ANC in AR glasses.

This paper presents a comprehensive analysis of performance for ANC in simulated AR glasses, using the AF method under free-field conditions with a noise source having variable direction-of-arrival (DOA).
The main contributions are:
(i) highlighting the need for a virtual microphone in ANC for AR glasses,
(ii) presenting the limitation of ANC with a fixed AF when noise-source direction changes, and
(iii) demonstrating the advantage of using AF bank to improve robustness to a change in noise-source DOA.

\section{ANC with an Auxiliary Filter}
\label{sec:AF}
\begin{figure}[t]
    \makebox[\columnwidth][l]{%
        \hspace*{-0mm}%
        \includegraphics[width=1.15\columnwidth, trim=0mm 0mm 8mm 4mm, clip]{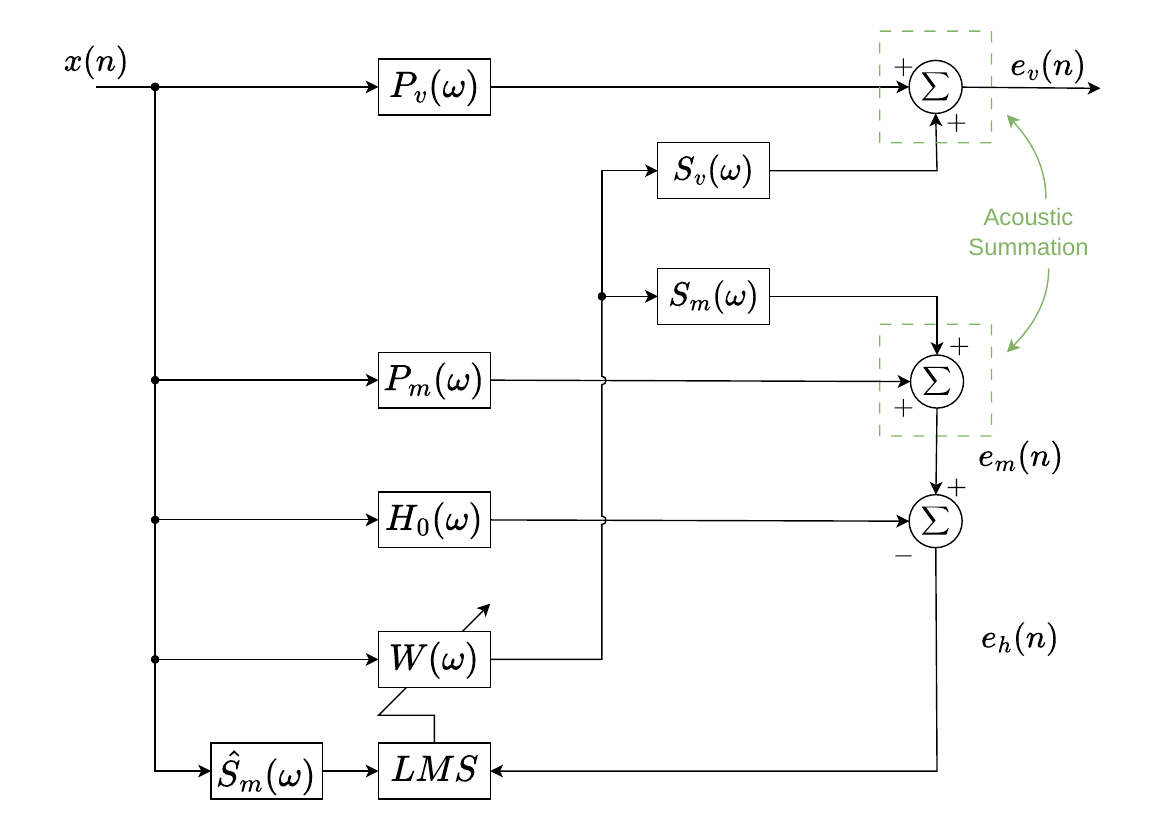}
    }
    \caption{Block diagram of the ANC system including the AF.}
    \label{fig:control}
\end{figure}

The ANC system assumed in this work is based on an AF and comprises a reference microphone, a monitoring microphone, a virtual microphone, and a control loudspeaker, and is based on \cite{shi_active_2020}. All microphones are used during the training stage, whereas during the control stage the virtual microphone is assumed to be inaccessible.
The virtual microphone signal in the frequency domain is given by
\begin{align}
    E_v(\omega) = P_v(\omega) X(\omega) + S_v(\omega) W(\omega) X(\omega),
    \label{eq:Ev}
\end{align}
where $X(\omega)$ is the reference signal; $P_v(\omega)$ and $S_v(\omega)$ denote the virtual primary and virtual secondary paths, respectively; $W(\omega)$ is the control filter; and $\omega$ denotes frequency.
The monitoring microphone signal $E_m(\omega)$ is defined analogously to \eqref{eq:Ev}, with $P_v(\omega)$ and $S_v(\omega)$ replaced by the monitoring primary and secondary paths, $P_m(\omega)$ and $S_m(\omega)$. During the control stage, the controller is adapted using a modified filtered-x least mean squares (FxLMS) algorithm \cite{kuo_active_1999}. The error signal used for adaptation, is given by
\begin{align}
    E_h(\omega) = E_m(\omega)-H(\omega)X(\omega),
\end{align}
as shown in Fig.~\ref{fig:control}, which presents a block diagram of the system. As in the FxLMS algorithm, the reference signal used for adaptation is prefiltered using an estimate of the monitoring secondary path $\hat{S}_m(\omega)$ (see Fig.~\ref{fig:control}).
Additional details on the AF method can be found in \cite{shi_active_2020, kajikawa_comparison_2019}. For the present analysis, the optimal AF in the frequency domain is given by
\begin{align}
    H_0(\omega) = P_m(\omega) - \frac{S_m(\omega)}{S_v(\omega)}P_v(\omega).
    \label{eq:H_0 paths}
\end{align}
With the optimal AF, $E_h(\omega)$ represents an error signal that tend to zero when the control filter converges to the optimal filter that minimizes the error at the virtual microphone, i.e.,
\begin{align}
    W_0(\omega) = -\frac{P_v(\omega)}{S_v(\omega)}.
\end{align}

\section{Theoretical Sensitivity Analysis of the Optimal AF}
\label{sec:Analysis}

\begin{figure}[t]
        \makebox[\columnwidth][l]{%
        \hspace*{-2mm}%
        \includegraphics[width=1.17\columnwidth, trim=0mm 10mm 10mm 12mm, clip]{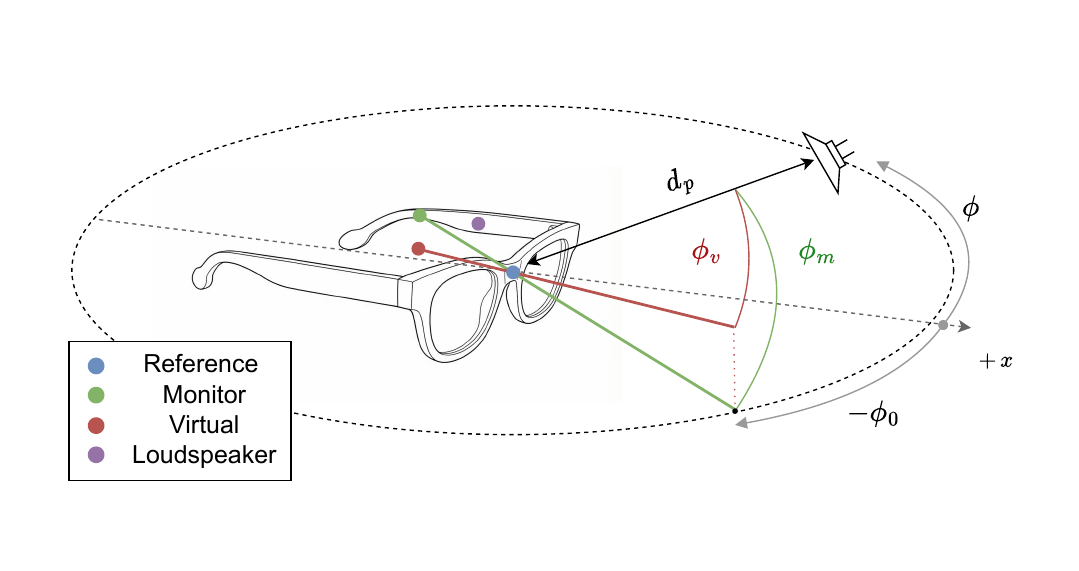}
    }
    \caption{AR glasses setup illustrating the positions of the microphones, along with the primary and secondary sources. This arrangement is used throughout the analysis and simulations.}
    \label{fig:analysis setup}
\end{figure}

The aim of this section is to develop a theoretical analysis that will assist in the understanding of the directional sensitivity of the AF. This will allow some prediction of the expected robustness of a fixed AF to changes in the direction of the primary noise source. To develop this theoretical insight, we consider an AR-glasses configuration equipped with microphones and a loudspeaker, arranged in accordance with the conventional AF system described in Section~\ref{sec:AF} and Fig.~\ref{fig:control}. A free-field acoustic environment is assumed for simplicity, in which the system is designed to attenuate noise at the wearer’s ear (the virtual microphone) originating from a single primary source with an arbitrary DOA. Both the primary and secondary sources are modeled as point sources; however, later in this section, the primary disturbance is approximated as a plane wave. Additionally, the primary noise is assumed to be observed at all microphones in the system, whereas the contribution of the secondary source at the reference microphone is neglected for simplicity. The coordinate system is centered at the reference microphone, with the glasses facing the positive $x$-axis such that the reference and monitor microphones lie in the $xy$-plane. Furthermore, the primary source is restricted to the $xy$-plane, where its DOA is given by the azimuthal angle $\phi$, defined from the positive x-axis toward the positive y-axis. The geometric arrangement used for the analysis is shown in Fig.~\ref{fig:analysis setup}.

Applying this model to the transfer functions in Eq.~\eqref{eq:H_0 paths}, we can express the optimal AF as
\begin{align}
    H_o(\omega, \phi)
    &= e^{-j \frac{\omega}{c} d_m \cos(\phi_m)} - G(\omega) e^{-j \frac{\omega}{c} d_v \cos(\phi_v)} 
    \label{eq:optimal AF}
\end{align}
where $\phi_m$ and $\phi_v$ denote primary source angles relative to the reference–monitor and reference–virtual axes, respectively (see Fig.~\ref{fig:analysis setup}), both of which are implicit functions of $\phi$; $d_m$ and $d_v$ are the distances separating the reference-monitor and reference-virtual microphone pairs, respectively; $G(\omega) \triangleq \frac{S_v(\omega)}{S_m(\omega)}$; and $c$ is the speed of sound.

Defining $\phi_0$ as the reference–monitor axis angle relative to the positive x-direction, see Fig.~\ref{fig:analysis setup}, we obtain $\phi_m = \phi - \phi_0$. Furthermore, we assume that the monitor microphone is placed on the glasses frame above the ear, as in the Project Aria design \cite{chime8_mmcs_aria}, such that $\phi_m \approx \phi_v$. Given this geometric framework, taking the derivative of the optimal AF in Eq.~\eqref{eq:optimal AF} with respect to the noise source DOA and collecting common terms yields:
\begin{align}
    \frac{\partial H_0}{\partial \phi} \approx & \,\, j \frac{\omega}{c} \sin(\phi-\phi_0) e^{-j \frac{\omega}{c} d_m \cos(\phi-\phi_0)} \notag \\
    & \times \Big[ d_m - G(\omega) d_v e^{-j \frac{\omega}{c}(d_v-d_m)\cos(\phi-\phi_0)} \Big]
\end{align}
Assuming that
\begin{align}
\frac{\omega}{c}\,\lvert d_v - d_m \rvert \ll 1,
\end{align}
i.e., that the difference between the distances is small compared to the wavelength, we obtain
\begin{align}
\frac{\partial H_0}{\partial \phi}
&\approx j\,\frac{\omega}{c}\,\sin(\phi-\phi_0)\,
e^{-j \frac{\omega}{c} d_m \cos(\phi-\phi_0)}
\Big[
d_m - G(\omega)\,d_v
\Big].
\end{align}
Accordingly, the magnitude of the derivative can be approximated as
\begin{align}
\left|\frac{\partial H_0}{\partial \phi}\right|
\approx |F(\omega)|\,|\sin(\phi-\phi_0)|,
\label{eq:dH approx}
\end{align}
where
\begin{align}
|F(\omega)|
= \frac{\omega}{c}\,\left|d_m - G(\omega)\,d_v\right|.
\end{align}
Equation~\eqref{eq:dH approx} clearly shows that the AF is more sensitive to changes in the primary source direction around $\phi=\phi_0 +90\degree$ compared to $\phi = \phi_0$.

\section{Experimental Study}
This section presents a simulation study conducted under simplified acoustic conditions aiming to: (i) demonstrate the necessity of the virtual microphone technique for ANC in AR glasses, (ii) evaluate the performance degradation caused by AF–DOA mismatch while demonstrating the associated theoretical sensitivity analysis, and (iii) show the advantage of using a finite AF bank during online control as a practical alternative to per-DOA AF calibration.

\subsection{Simulation Setup}
\label{sec:setup}
\begin{figure}[b]
    \centering
    \includegraphics[width=\columnwidth, trim=0mm 20mm 0mm 10mm, clip]{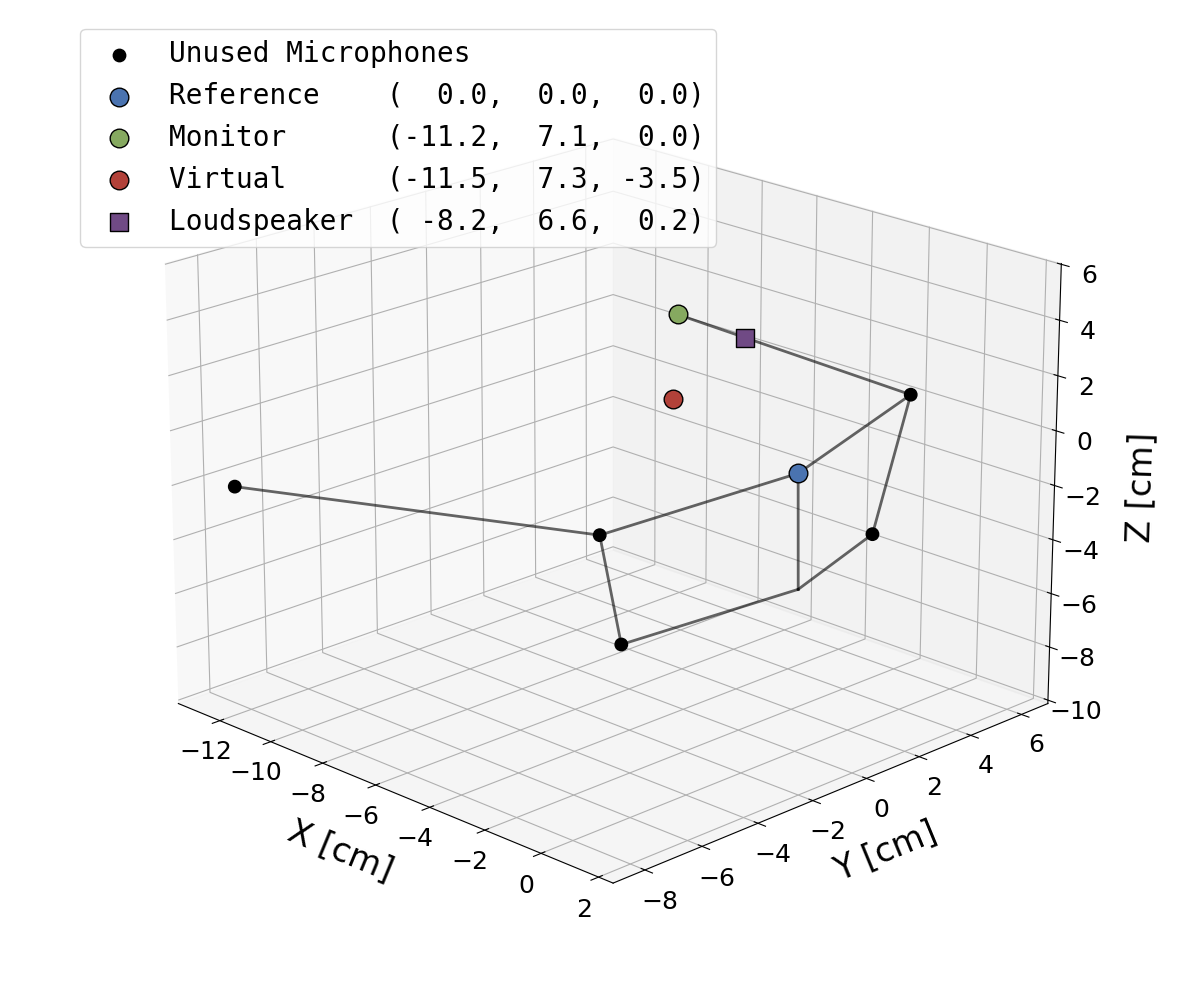}
    \caption{Geometric layout based on the Project Aria glasses geometry\cite{chime8_mmcs_aria}, including the loudspeaker and virtual microphone. Coordinates are provided in the legend in centimeters.}
    \label{fig:aria}
\end{figure}

The microphone positions used in the simulations are taken from the Project Aria specifications \cite{chime8_mmcs_aria}. For visualization purposes, all Aria microphones are shown in Fig.~\ref{fig:aria}, although only a subset is used in the simulation, as denoted by the colored marks on the figure. A loudspeaker, serving as the secondary source, is assumed to be mounted on the frame, and a virtual microphone is positioned at the ear location.

A free-field environment is assumed, and both the primary and secondary sources are modeled as point sources. Simulations are performed at a sampling rate of $24\,\mathrm{kHz}$, and no additional processing or hardware delay is included.

For each source azimuth $\phi$ in the $x$-$y$ plane (fixed elevation), we simulate a $5\,\mathrm{s}$ realization where the noise source is positioned at a distance of $d_p = 3\,\mathrm{m}$ from the reference microphone (see Fig.~\ref{fig:analysis setup}). This distance allows the simulated point source to reasonably approximate the plane wave behavior assumed in the theoretical analysis. The primary disturbance is generated as Gaussian white noise, band-limited to $100$–$800\,\mathrm{Hz}$ using a 4th-order Butterworth filter, reflecting the low-frequency range where ANC is most effective.

\subsection{Methodology and Evaluation}
Simulations are conducted using the normalized FxLMS (NFxLMS) \cite{kuo_active_1999} and AF-based ANC frameworks. Following the procedure in \cite{shi_active_2020}, the AF is calibrated using the LMS algorithm and the noise signal described in~\ref{sec:setup} before being fixed for the control stage. All adaptive filters are finite impulse response filters with 256 coefficients and are initialized with a value of zero. This length provides a sufficient time window to model the relevant acoustic paths at the 24 kHz sampling rate. For simplicity, the secondary path estimate is assumed to be perfectly known. For each control filter adaptation, the reported attenuation corresponds to the best performance obtained over the step sizes $\mu \in \{10^{-4},\,10^{-3},\,10^{-2},\,10^{-1}\}$. The performance metric used throughout the simulations is the mean-squared attenuation in dB, computed from the disturbance and error signals of the virtual microphone over the last 15\% of each $5\,\mathrm{s}$ realization to reflect steady-state convergence.

\subsection{Necessity of the Virtual Microphone}
\begin{figure}
    \centering
    \includegraphics[width=\columnwidth, trim=0mm 5mm 0mm 5mm, clip]{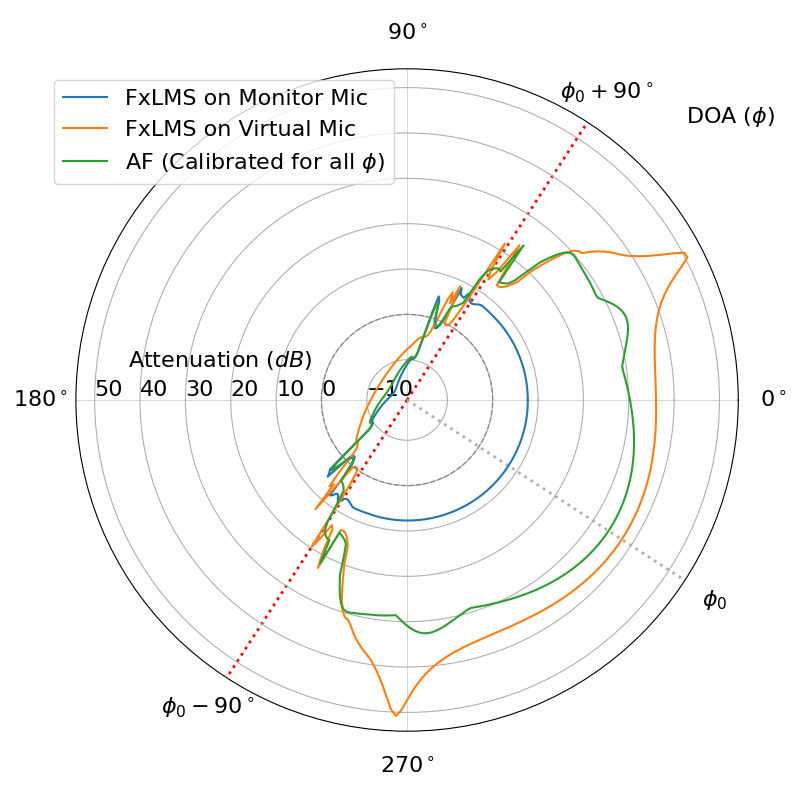}
    \caption{Polar plot of attenuation at the virtual microphone (at the ear) versus source azimuth. Comparison between NFxLMS control using the monitoring microphone, NFxLMS control using the virtual microphone, and AF control calibrated separately for each azimuth.}
    \label{fig:neccessity}
\end{figure}

To investigate the necessity of the virtual microphone technique, we compare the attenuation at the virtual microphone achieved by three configurations: (i) monitor-based NFxLMS (where adaptation is driven by the monitor microphone signal), (ii) virtual-based NFxLMS (the "ground truth" case with adaptation directly at the virtual microphone), and (iii) the AF method calibrated separately for each azimuth. Figure ~\ref{fig:neccessity} displays the attenuation results as a polar plot, where the angular coordinate corresponds to the noise source DOA. This visualization format is utilized for all subsequent experiments.
The disparity in performance between the monitor-based NFxLMS and the virtual NFxLMS highlights the necessity of implementing a virtual microphone technique. The AF method proved effective in bridging this gap, achieving attenuation levels that approach the performance of the virtual NFxLMS. Consequently, the AF method presents a promising potential solution for practical implementation.

The attenuation exhibits approximate symmetry around $\phi \approx \phi_0 \approx 327^\circ$, corresponding to the direction of the reference–monitor axis ($\phi_0$). At this azimuth, the relative propagation delay between the reference and the monitor/virtual microphones is maximal, resulting in favorable causality conditions for control. When the azimuth deviation exceeds approximately $\pm 90^\circ$, the noise reaches the monitor/virtual microphone before the reference, requiring a noncausal controller. In this region, all algorithms degrade significantly.

\subsection{AF sensitivity to DOA mismatch}
\begin{figure}
    \centering
    \includegraphics[width=\columnwidth, trim=0mm 5mm 0mm 5mm, clip]{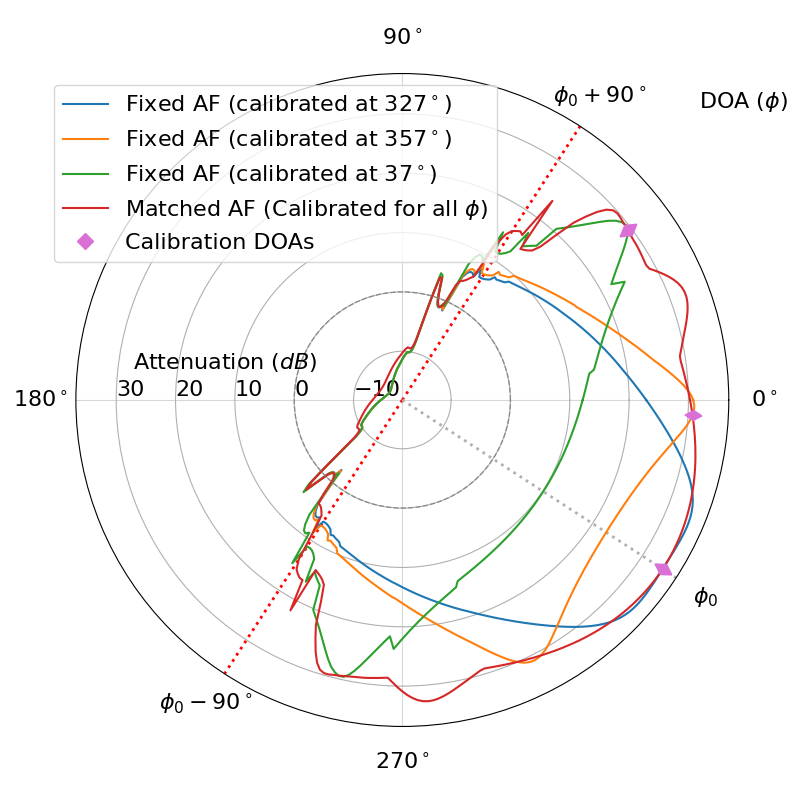}
    \caption{Polar plot of attenuation at the virtual microphone (at the ear) versus source azimuth. Comparison between AF control calibrated separately for each $\phi$ and AF control using a single AF calibrated at $\phi_0$ or $\phi_0 +30^\circ$ or $\phi_0 +70^\circ$.}
    \label{fig:single AFs}
\end{figure}

We evaluate the performance degradation caused by AF mismatch between calibration and control, by applying an AF calibrated at a single DOA of the primary source, to other source DOAs. As shown in Fig.~\ref{fig:single AFs}, the angular region around the calibration DOA in which the fixed AF achieves performance comparable to the fully calibrated AF is wider for the AF calibrated at $\phi_0$ than for the one calibrated at $\phi_0 + 60^\circ$. This indicates that the AF exhibits greater robustness to DOA mismatch in the angular region of the reference–monitor axis than at more distant azimuths.

This behavior is consistent with the analysis presented in section ~\ref{sec:Analysis}. In the analysis, the sensitivity term is proportional to $|\sin(\phi-\phi_0)|$, which attains its minimum at $\phi=\phi_0$, (the reference–monitor axis). As the noise source azimuth deviates from this axis within the range $[\phi_0 -90^\circ , \phi_0 + 90^\circ]$, the sensitivity of the optimal AF increases, implying that such a DOA mismatch may induce a larger variation in the AF.

\subsection{Finite AF Bank}
To provide a practical alternative to per-DOA AF calibration, we examine the performance obtained when selecting the AF from a finite bank based on the primary source direction. Given the significant performance degradation observed in the non-causal regions (where the azimuth deviation exceeds $\pm 90^\circ$ relative to $\phi_0$), the AF bank is constructed by calibrating AFs at uniformly spaced DOAs within $[\phi_0 -90^\circ, \phi_0 + 90^\circ]$. For each azimuth, the AF is selected from the bank according to the calibration DOA closest to the  noise source direction.

\begin{figure}
    \centering
    \includegraphics[width=1\columnwidth]{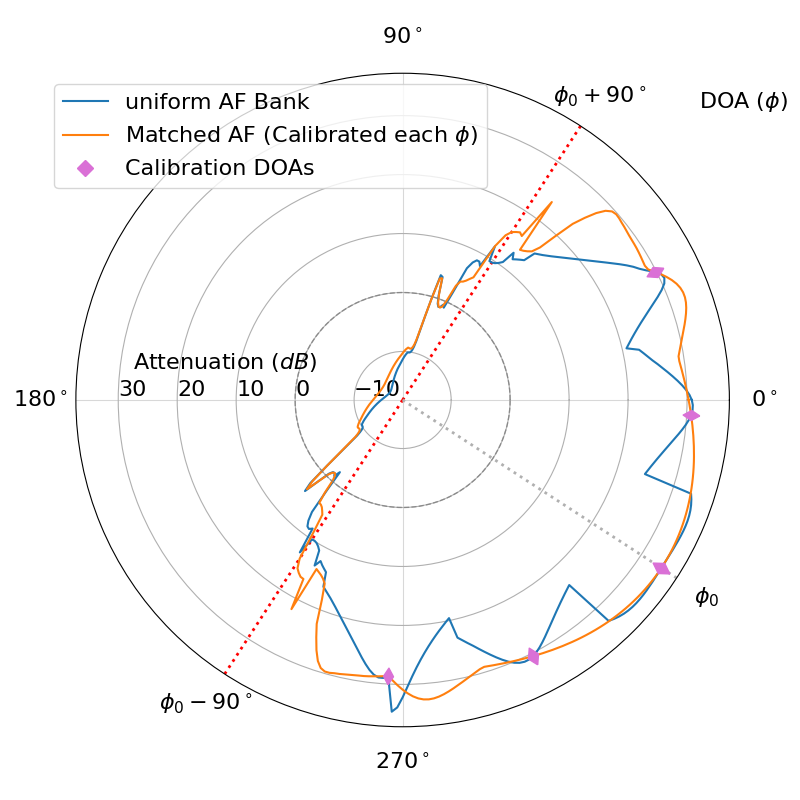}
    \caption{Polar plot of attenuation at the virtual microphone (at the ear) versus source azimuth. Performance obtained when the AF is selected from a finite AF bank of size 5, compared to fully calibrated AF control for each $\phi$.}
    \label{fig:bank}
\end{figure}

As observed in Fig.~\ref{fig:bank}, selecting the AF from a finite bank significantly improves the performance compared to using a single fixed AF, while still not fully matching the performance achieved with per-azimuth calibration. To quantify how closely the finite bank performance approaches the fully calibrated case, Table~\ref{tab:af_bank} compares the achieved attenuation averaged over the azimuth range $\phi_0 \pm 90^\circ$ (i.e., where the controller operates under causal conditions), for different bank sizes.

\begin{table}[h]
\centering
\caption{Attenuation in dB at the virtual microphone, averaged over primary source DOA at the causal azimuth range $\phi_0 \pm 90^\circ$ for different AF bank sizes compared to fully calibrated AF control.}
\label{tab:af_bank}
\begin{tabular}{c c c}
\hline
AF Bank Size & Mean Attenuation [dB] & Gap to Fully Calibrated [dB] \\
\hline
3 & 22.7 & 6.3 \\
4 & 24.0 & 5.0 \\
5 & 25.3 & 3.7 \\
6 & 26.0 & 3.0 \\
7 & 26.5 & 2.5 \\
Fully calibrated & 29.1 & 0 \\
\hline
\end{tabular}
\end{table}

\section{Conclusions}
This study validates the necessity of the virtual microphone technique for effective ANC in AR glasses. We further established a theoretical framework that characterizes the performance degradation caused by the mismatch between calibration and control due to a change in the noise arrival directions. To mitigate this sensitivity, we proposed an AF bank strategy; simulations demonstrate that using a bank of just 5 filters leads to only a small degradation in averaged performance, offering a robust practical alternative to full recalibration. While this work focused on a single-channel implementation with a single noise source in free-field conditions, future research is proposed to extend this evaluation to more realistic conditions including reverberation, multiple noise sources, and multichannel control strategies.

\bibliographystyle{IEEEtran}
\bibliography{references}

\end{document}